\documentstyle[12pt,aasms]{article}
\tightenlines

\begin{document}

\begin{center}
Accepted for publicaton in {\it The Astronomical Journal}
\end{center}

\vskip 24pt

\title{The Extinction Law in an Occulting Galaxy}

\vskip 24pt

\author{
Andreas A. Berlind, 
A. C. Quillen \altaffilmark{1},
R. W. Pogge,
K. Sellgren \altaffilmark{2}}

\bigskip
\bigskip

\affil{Department of Astronomy, The Ohio State University, 174 W. 18th Ave., 
\\Columbus, OH 43210-1106}
\affil{Electronic mail: aberlind,pogge,sellgren@payne.mps.ohio-state.edu
\\aquillen@as.arizona.edu}
\altaffiltext{1}{Current address: Steward Observatory, University of Arizona, 933 N. Cherry Ave., Tucson, AZ 85721}
\altaffiltext{2}{Visiting Astronomer, Cerro Tololo Inter-American Observatories.}

\clearpage

\centerline
{\bf Abstract}                   

We measure the extinction law in a galaxy's spiral arm and interarm
regions using a visual and infrared ({\it BVRJHK}) imaging study 
of the interacting galaxies NGC 2207 and IC 2163.  This is an
overlapping spiral galaxy pair in which NGC 2207 partially occults 
IC 2163.  This geometry enables us to directly measure the extinction 
of light from the background galaxy as it passes through the
disk of the foreground galaxy.  We measure the extinction
as a function of wavelength, and find that there is less extinction
in the optical bands than expected from a normal Galactic extinction law.  
This deviation is significantly larger in the interarm region 
than in the spiral arm.  The extinction curve in the spiral arm
resembles a Milky Way $R_V=5.0$ dust model and the interarm extinction
curve is flatter (``greyer'') still.  We examine the effect of scattering
of background galaxy light into the line of sight and find that it is 
negligible.  We also examine the effect of an unresolved patchy dust 
distribution using a simple two-component dust model as well as the clumpy dust
model of \cite{wit96}.  Both models clearly demonstrate that an unresolved 
patchy dust distribution can flatten the extinction curve significantly.  
When fit to the data, both models suggest that the observed difference between 
the arm and interarm extinction curves is caused by the interarm region of 
NGC 2207 having a higher degree of dust patchiness (density ratio between 
high-density and low-density phases) than the spiral arm region.
We note that an unresolved patchy dust distribution will cause us to
underestimate the average column depth of gas in a galaxy if based solely
on the visual extinction.  It is much better to use the infrared extinction
for this purpose. 


\clearpage

\section {INTRODUCTION}

Little is known about the extinction law for the large-scale dust distribution
in spiral galaxies.  There is even some controversy as to how large the total
extinction through a spiral galaxy is.  \cite{val90} presented evidence that 
spiral disks are opaque.  \cite{bur91} pointed out selection effects in 
Valentijn's study; nevertheless, they also found that galaxies are optically 
thick.  On the other hand, these results are in direct conflict with the 
simple fact that we can see out of the disk of our galaxy.  

A more direct approach to solving this problem has been taken by 
\cite{whi92}.  They proposed that the optical depth in a spiral galaxy may be
measured directly if it lies in front of another object.  The ideal case
consists of a face on symmetric spiral galaxy half overlapping with a 
symmetric, background spiral or elliptical galaxy.  The galaxies need
to be symmetric so that their non-overlapping parts may be used to estimate
the surface brightness of their overlapping parts.  
\cite{whi92} obtained {\it B} and {\it I} images of such a candidate, AM1316-241,
and found that the observed extinction is largely confined to the spiral
arm regions, leaving the interarm regions mostly transparent.  
Most recently, \cite{whi96} have studied 10 overlapping galaxy pairs and 
reached a similar conclusion in all cases.
The advantages of this direct approach to studying the opacity of spiral disks 
are numerous and are described in detail by \cite{whi96}.  
Because the background galaxy is behind the foreground 
galaxy, the approximation of the foreground galaxy as a dust screen is a good 
one.  This geometry is much simpler than that in a normal disk galaxy 
where dust and stars are mixed.  

Previous studies of occulting galaxies have been restricted to visual 
wavelengths.  
In this paper we perform a similar study of the extinction through the disk of
a spiral galaxy, extending our wavelength base into the near infrared.
We present a six band ({\it BVRJHK}) study of the interacting galaxy pair 
NGC 2207 and IC 2163.  This pair consists of two symmetric, almost face-on 
interacting spiral galaxies which are partially overlapping.  The galaxies' Hubble 
types are SAB(rs)bc pec and SB(rs)c pec, for NGC 2207 and IC 2163, respectively 
(\cite{dev91}).   
These galaxies have been observed and modeled in detail by \cite{elm95a} and
\cite{elm95b}, whose HI and optical observations indicate that the 
two galaxies have probably gone through a recent nonmerging encounter.   
\cite{elm95a} do not find any significant excess of star 
formation in these galaxies as a result of their interaction.

By determining the extinction through the foreground galaxy disk at many 
wavelengths, in both spiral arm and interarm regions, we can sample the 
effective large-scale extinction curve, and compare it to Galactic 
dust models.  In sec. 3 we describe how we derive the extinction curve.
In sec. 4 we present our results and discuss deviations of our observed
extinction curve from Galactic dust models.

\section {OBSERVATIONS} 

NGC 2207 and IC 2163 are good candidate galaxies for the following reasons:
1) both galaxies display good symmetry, without which we cannot estimate
their intrinsic colors in the overlapping region;
2) The foreground galaxy is close to face on; this geometry provides a small 
optical path through the foreground disk and facilitates our estimation
of scattering effects since the optical properties of the disk are 
dependent on inclination with increasing sensitivity at high inclinations
(\cite{bru88}); 3) the two galaxies are positioned so that
they partially overlap, but still have large enough non-overlapping
regions to allow estimates of their true colors; 
4) The simulations of \cite{elm95b} supply us with a distance between the 
two galaxies which is necessary for determining the effect of scattered 
light on our extinction estimates.

These galaxies were observed as part of an imaging survey being carried
out by the Ohio State University of $\sim 200$
galaxies.  The survey's goal is to produce a library of photometrically 
calibrated images of late-type galaxies from $0.4$ to $2.2 \mu$m.
For notes on the observation and reduction techniques see
\cite{fro97}, or for individual examples, see \cite{qui94}, and \cite{qui95}.
All the images were obtained at CTIO on the 1.5m telescope.
The {\it BVR} images were observed on 1995 March 7 using the Tek\#2 
$1024^2$ pixel CCD with a spatial scale of $0.\hskip-2pt''44$/pixel.
Total on-source exposure times were 20min, 15min and 10min for {\it B,V,} and 
{\it R} respectively.  The {\it JHK} images were 
observed on 1995 March 14 using the NICMOS 3 $256^2$
pixel infrared array with a spatial scale of $1.\hskip-2pt''16$/pixel.
Total on-source exposure times were 16min, 15min and 29min for {\it J,H,} 
and {\it K} respectively. 

A color image was constructed and is displayed as a color plate 
in Figure 1.
The color image was composed by combining the {\it B,R,} and {\it H}-band
images, coding them as Blue, Green, and Red intensities, respectively.
The relative scaling between the images was chosen so as to neutralize
the colors of most field stars, which serves to emphasize dusty regions
as reddish and regions of star formation as blue, while the intensities
preserve the general photometric appearance of the galaxies.

It is immediately noticeable (see Fig. 1) that the background spiral 
galaxy can be easily seen through the foreground galaxy even though a 
large portion of it lies directly behind one of the foreground galaxy's 
spiral arms.  Moreover, reddening of the background galaxy is plainly visible 
in this region, and the dust content of the foreground galaxy, which 
is outlined by the reddening, appears to be closely concentrated in its 
spiral arms.
In Figure 2 we present {\it B} and {\it H} greyscale images of the two 
galaxies.  The {\it B} image shows that the foreground dust lane causes
significant extinction of the background galaxy light, whereas in the {\it H} 
image the extinction is minimal, as expected. 

\section {THE EXTINCTION CURVE}
\subsection{Measuring the Extinction in Each Band}

To calculate the extinction through the disk of the foreground galaxy 
we use the method pioneered by \cite{whi92}.  
We call the surface brightness of the foreground spiral $F$, and that of the
background spiral $B$.  The observed surface brightness of the overlapping 
region will then be
\begin{equation}
F+Be^{-\tau}
\end{equation}
where $\tau$ is the optical depth through the foreground spiral.  The steps 
we follow to measure $\tau$ for each band are as follows:
a) we estimate $B$ for regions where the two galaxies
overlap, based on $B$ in non-overlapping regions (more detail given in 
sec. 3.2); this estimate requires that the galaxy is symmetric;
b) we estimate $F$ by following a similar procedure as in step (a);
c) we measure the surface brightness in the overlapping regions;
d) we use eq. (1) to determine $\tau$ as a function of position in the
overlapping region of the foreground galaxy.  More specifically, we
measure the extinction $A_{\lambda}$ (where $A_{\lambda} = 1.086\tau_{\lambda}$)
in the spiral arm and interarm regions.  

\subsection{Spiral Arm and Interarm Regions}

We measure the surface brightness in square regions approximately
$3.\hskip-2pt''5$ wide ($1''= 170$pc for a distance to the galaxy of 35Mpc),
taking the mean pixel value to be our surface brightness estimate, and
the uncertainty in the mean to be the error in our measurement.
We do this for five different regions located along the foreground galaxy's 
outer spiral arm, where it occults the background galaxy.
These five regions, which are located at a radius $\sim 1.\hskip-2pt'3$ from 
the center of the foreground galaxy, are marked in Figure 2.  We then 
estimate the background galaxy's true surface brightness by measuring it 
in five new regions, each of which is located at an opposite location 
(symmetrically reflected about the galaxy nucleus) from its parent region.
Finally, we follow a similar procedure for the foreground galaxy to estimate 
its contribution to the total light seen in each of the overlapping regions.  
Due to the absence of a good counterpart for the foreground occulting arm,
we estimate its average surface brightness from its own non-overlapping
part.  In this way, we obtain five estimates of the extinction through the 
foreground galaxy's spiral arm.  Our final extinction estimate is the average 
of these five measurements.  The errors are estimated by propagating
the errors in each step of the above process.

To measure the extinction in the interarm region,
we repeat the same procedure, this time placing our first five squares in the 
interarm region just inside the foreground galaxy's outer spiral arm.  These
boxes, which are located at a radius $\sim 1.\hskip-2pt'1$ from the galaxy's 
center, are also marked in Figure 2.
For comparison with the sample of occulting galaxies presented by \cite{whi96},
the $B=25$mag/arcsec$^2$ isophotal radius of NGC 2207 is 
$R_{25}^{{\it B}} \sim 1.\hskip-2pt'7$.
In these units, the regions that we examine are $R/R_{25}^{{\it B}} = 0.76$
for the spiral arm region and $R/R_{25}^{{\it B}} = 0.65$ for the interarm 
region.

Although our extinction measurements are independent of absolute 
calibration, we can check the quality of our galaxy color estimates by 
comparing them to other galaxy colors.
The estimated colors of the foreground and background galaxies in the 
overlapping arm and interarm regions are listed in Table 1.
Comparing these to galaxy colors found by \cite{fro88}, we are reassured 
that they are reasonable colors for Sbc and Sc spiral galaxies.

A good discussion of errors is given by \cite{whi96}.  The greatest
source of error in our extinction measurements is possible asymmetries in the 
background galaxy.  Since the foreground galaxy is substantially fainter than 
the background galaxy, asymmetries in the foreground galaxy will not greatly 
affect our estimates of $A_{\lambda}$.  Small errors may also be 
introduced by imperfections in the flat-fielding and sky-subtraction of the 
images; however, these errors will not be as large as those caused by asymmetries
in the background spiral galaxy.  We estimate the average error in $A_{\lambda}$
that could be caused by such asymmetries by examining the background galaxy's
deviations from symmetry in its non-overlapping regions.  We find that the 
error in the measured extinction ranges from $\Delta A_{\lambda} \sim 0.08$, 
in {\it B}, to $\Delta A_{\lambda} \sim 0.03$, in {\it K}.  The errors are 
smaller in the infrared bands because the background galaxy is much smoother 
(and, therefore, more symmetric) in the infrared than it is in the visible. 
The photometric errors, on the other hand, range from $\Delta A_{\lambda}, 
\sim 0.06$ in {\it B}, to $\Delta A_{\lambda} \sim 0.03$, in {\it K}.  We sum 
these two errors in quadrature to obtain total error estimates for the 
extinctions we measure.   

\section {RESULTS}

Our measurements for the {\it BVRJHK} extinction through NGC 2207 in its spiral
arm and interarm regions are listed, along with their uncertainties, in
Table 2; the total extinction errors that we find range from 0.1mag, in {\it B}, 
to 0.04mag, in {\it K}.  Like \cite{whi92}, we find that the extinction is mainly
concentrated in spiral arms (where $A_{\it V} \sim 1.0$), while the interarm 
regions (where $A_{\it V} \sim 0.5$) are mostly transparent.  NGC 2207 has 
values for $A_{\it B}$ at $R/R_{25}^{{\it B}}$ similar to one of the dustier 
galaxies studied by \cite{whi96}. 

Also listed in Table 2 is the ratio $F/B$ of the foreground galaxy surface 
brightness to that of the background galaxy.  This ratio indicates that 
the foreground galaxy's contribution to the total light in the regions
examined is small.  Our extinctions will, therefore, not be greatly 
affected by errors in our estimate of $F$.
   
In Figure 3 we plot the extinction, normalized to the extinction in the J band,
as a function of band wavelength.  We are surprised to find that the 
extinction curve for the spiral arm region seems ``flatter'' (``greyer'')
than expected from a standard $R_V=3.1$ dust model, where $R_V=A_V/E(B-V)$.  In 
other words, the visual extinction is found to be less than expected 
compared to that in the near-infrared.  The extinction curve appears to 
agree with an $R_V=5.0$ dust model.  This effect is even more pronounced in 
the interarm region extinction curve, where even an $R_V=5.0$ dust model 
overestimates the visual extinction.

For the bulk of the ISM in the Milky Way, the observed extinction law is 
$R_V=3.1$; values of $R_V=5.0$ are found in molecular clouds and HII regions
(\cite{sav79}; \cite{car89}).
Deviations from this law, such as the one that we find, have been 
detected in various systems.  One such system is the Sombrero galaxy, an edge-on
system studied by \cite{ems95} where stars and dust are mixed.  Although
\cite{ems95} also observed an $R_V=5.0$ extinction law, the geometry of the
Sombrero galaxy is extremely different from that of our overlapping galaxies.
Also, \cite{bru88} modeled the effects of dust mixed with stars and found
that some deviations from an $R_V=3.1$ law are expected due to scattering of
light.  \cite{whi96} calculated $A_B/A_I$ for their overlapping galaxy 
candidates and found that some of them seem to have a flatter extinction 
curve than the Milky Way.  Our more detailed study of the extinction 
curve confirms this finding.  

Two effects may be responsible for the high $R_V$ for the extinction curve
that we observe.  Scattering of light from the background galaxy into the 
line of sight can cause us to overestimate the amount of light successfully
passing through the foreground disk and, thus, underestimate the 
extinction.  This effect will be greater in the visual bands where
scattering is more efficient, so a flattening of the extinction curve is 
expected.  An unresolved patchy dust distribution also could be responsible 
for the high value of $R_V$.

\subsection{Scattering of Background Galaxy Light into the Line of Sight}

We can estimate the maximum effects of scattering by adopting dust grain 
properties and considering the geometry of our galaxies.  
We assume that the geometry 
consists of a simple dust screen (NGC 2207), perpendicular to
our line of sight, that partially occults the background face-on galaxy
(IC 2163).  The scattering effects will be maximum if we consider the 
background galaxy to be a point source having a small scattering angle, $\theta$.
In this case, the maximum surface brightness in each band 
that could be scattered into the line of sight is approximated by 
\begin{equation}
S_{max} = \frac{\tau \omega B_{tot} H(\theta)}{4\pi} \left( \frac{D}{a} \right)^2 (4.85\times10^{-6})^2
\end{equation}   
where $\tau$ is the optical depth, $\omega$ is the 
dust albedo, $B_{tot}$ is the total brightness of the background galaxy, 
$D$ is the distance to the background galaxy, $a$ is the 
distance between the two galaxies, and last numerical factor converts $S_{max}$
into units of flux/arcsec$^2$.  Here we have assumed the Henyey-Greenstein 
phase function $H(\theta)$ which, for $\theta=0$, is
\begin{equation}
H(0) = (1-g^2)(1+g^2-2g)^{-3/2} 
\end{equation}   
where $g$ is the phase function asymmetry (we adopt $\theta=0$ because this
maximizes $S_{max}$).

Using N-body simulations and a distance to the pair of $35$Mpc, \cite{elm95b} 
found that $a = 40$Kpc.  So for our system
\begin{equation}
S_{max} = 1.43\times10^{-6} \tau \omega B_{tot} \frac{1+g}{(1-g)^2}.
\end{equation}
There is some uncertainty as to the values of $\omega$ and $g$ for the 
near-infrared (especially at {\it K}).  Many studies have suggested that 
the dust albedo and phase function asymmetry in the near-infrared 
are much higher than the predicted values of \cite{dra84} (e.g., \cite{joh90}; 
\cite{pen90}; \cite{sel92}; \cite{wit94}; \cite{shu95}; \cite{leh96}).  In order 
to ensure that we are calculating the maximum effect
of scattering, we use the high values for $\omega$ and $g$ chosen by 
\cite{wit96}.  These values are $\omega = 0.6$ and $g=0.6$, in all bands.

Table 3 lists the values we use to estimate $S_{max}$, the percent 
contribution of scattered light to the total light we see, and the maximum
effect of scattering on our extinction measurements.
Our values for $\tau$ may be found in Table 2 (where 
$\tau_{\lambda} = 0.921A_{\lambda}$).  
We estimate $B_{tot}$ by integrating all the light from
the right half of the background galaxy (the half which is not occulted by the
foreground galaxy) and multiplying by two.  Table 3 
lists our values for $B_{tot}$ and $S_{max}$ in all observed bands.  
Table 3 also lists the value of the ratio $S_{max}/(F+Be^{-\tau})$
which indicates what fraction of the light we see in overlapping regions
could be scattered light.  Finally, Table 3 lists 
$\Delta A_{\lambda}/A_{\lambda}$, the maximum errors in our extinction estimates 
that could result from scattering of light.  

In all bands, $\Delta A_{\lambda}/A_{\lambda}$ is less than 0.033.  
For purpose of comparison, $\Delta A_{\lambda}/A_{\lambda}$ due to photometric
and asymmetry uncertainties is greater than 0.073 in all bands, i.e., the
combined photometric and asymmetry uncertainties (listed in Table 2)
are much greater than the effects of scattering.  Clearly, scattering
cannot account for the high $R_V$ value that we observe.  
Moreover, the effect of scattering on the interarm extinction curve is
smaller than that on the spiral arm extinction curve, thus, making it even 
harder to explain their significant difference as due to scattering.

\subsection{Patchy Dust Distribution in NGC 2207}

Since scattering effects cannot explain the deviations from
the Galactic extinction curve that we observe, we examine the possibility 
that this deviation is caused by
unresolved clumping of the dust.  It is reasonable to expect a patchy dust 
distribution in NGC 2207.  The dust distribution in our own galaxy is 
patchy on scales small enough that the patches would be 
unresolved if seen from a distance of $35$Mpc (\cite{sav79}).  
Moreover, \cite{rix93} 
have mapped the dust distribution in M31 and have shown that it is 
quite patchy in appearance.  \cite{whi96} suggest that clumpiness of dust
is probably the cause for their overlapping galaxies' flat extinction curves.
\cite{wit96} have studied the radiation transfer in a two-phase
clumpy dust medium and have also suggested that a clumpy medium tends to flatten 
the effective extinction curve. 

Is a patchy dust distribution in NGC 2207 responsible for flattening
our derived extinction curve?  
To investigate this we explore a simple two-dimensional, two-component dust model.
We assume that the dust in the foreground galaxy consists of a high density 
component and a low density component.  The high
density dust component has a {\it J}-band optical depth $\tau_1$, and 
an area filling factor $f$.  The low density component has a 
{\it J}-band optical depth $\tau_2$
and an area filling factor $1-f$.  The effective transmission through the
patchy dust distribution will then be
\begin{equation}
e^{-\tau_{eff}} = fe^{-\tau_1} + (1-f)e^{-\tau_2}
\end{equation}
where $\tau_{eff}$ is the effective optical depth that we measure.
We assume that each dust component follows an $R_V=3.1$ Galactic extinction 
law, and study the extinction curve for $\tau_{eff}$.  We find that any 
combination of values for our free parameters $\tau_1$, $\tau_2$, and 
$f$ tends to flatten the effective extinction curve.  
In particular, the slope of the effective extinction curve appears to be
most sensitive to differences in $\tau_1 / \tau_2$ and $f$;
the curve is flatter for larger values of $\tau_1 / \tau_2$ and for values 
of $f$ which are closer to $50\%$.

For purposes of illustration, we choose two sets of parameters, one for the 
spiral arm and one for the interarm, that reproduce the effective extinctions 
that we measure (and thus, the extinction curve that we observe.)
These parameters are: $\tau_1 = 2.0$, $\tau_2 = 0.22$ and $f = 0.12$ 
for the spiral arm region, and $\tau_1 = 2.0$, $\tau_2 = 0.08$ and $f = 0.15$ 
for the interarm region.  These models give effective {\it J}-band extinctions of
$A_{eff}=0.35$ and $A_{eff}=0.22$, respectively, which match the measured
extinctions in Table 2.  We plot these models in Figure 4 to illustrate 
that the two-component dust model is capable of reproducing our extinction
curves.  We are cautious about assigning any physical significance to these
specific models however, because their uniqueness is not established.

In order to fully explore the parameter space of our two-component model,
we compute $\chi^2$ for our measured extinctions, keeping $f$ fixed
and varying $\tau_1$ and $\tau_2$.  In Figure 5, we show contours of constant
reduced-$\chi^2$ for four different values of the filling factor $f$.  
Contours are shown for both the spiral arm (bold) and the interarm (light) 
extinctions.  We note that $\tau_2$ is reasonably tightly constrained
and does not show a significant dependence on $\tau_1$ or $f$.  $\tau_1$, on the
other hand, is not that well constrained.
The ``golf club'' contours in Figure 5 reveal the main difference 
between the spiral arm and interarm regions: the low density dust component,
$\tau_2$, is smaller in the interarm region than it is in the spiral arm 
region.  In other words, the observed arm-interarm difference can be explained
by a model containing high density dust clumps embedded in a low density
interclump medium which is denser in the galaxy's spiral arms, and less dense in 
its interarm regions.  An example of this model would be molecular 
clouds (corresponding to $\tau_1$) embedded in a lower density dusty environment
(corresponding to $\tau_2$).  One could then speculate that, as the ISM moves
from an interarm into a spiral arm region, the low density interclump regions are 
compressed ($\tau_2$ increases), whereas the molecular clouds, being bound 
systems, are not much affected ($\tau_1$ stays approximately the same).  
According to the two-component patchy dust model, this scenario may be able to
explain the extinction laws that we observe.

In addition to our two-component patchy dust model, we also explore the clumpy
dust model of \cite{wit96}.  This is a three-dimensional two-phase model.
The model consists of a spherical volume divided into $N^3$ bins, each of which 
is randomly assigned to either a high-density state or a low-density state.
Photons originate from a centrally located, isotropically radiating point 
source and their transfer through the clumpy medium is studied.  
Parameters of this model are $N$, the volume filling factor $ff$, the density
ratio between the low-density and high-density phases $k_2/k_1$, and the optical
depth of the equivalent homogeneous dust distribution $\tau_H$.  The model
is described in great detail by \cite{wit96}.  
The main advantage of this model over our two-component model is that it is 
three-dimensional, thus allowing for each line of sight to have a variety 
of optical depths (depending on the number of high-density bins that are 
encountered).  Although its geometry differs from that of our overlapping galaxy 
pair (i.e., it contains a point source of light embedded in a spherical dust 
distribution, rather than an extended light source behind a flat dustscreen), 
the \cite{wit96} model is, in this case, essentially indistinguishable from a 
similar model of planar geometry since scattering is negligible.

We fit this clumpy dust model to our measured visual and infrared extinctions 
by minimizing $\chi^2$.  We perform the fits for the case $N=10$, and our free 
parameters are $k_2/k_1$ and the {\it J}-band value of $\tau_H$.  The fits are 
relatively crude because we do not explore the full parameter space of the model; 
nevertheless, they suffice to demonstrate that the \cite{wit96} clumpy dust model 
is capable of producing our observed extinction curves.  
In Figure 6 we present the best fits for the spiral arm and interarm extinction
curves.  The best-fit values of our parameters are $k_2/k_1=0.032$, $\tau_H=0.58$
for the spiral arm region, and $k_2/k_1=0.0032$, $\tau_H=0.50$ for the interarm 
region.  These models give effective {\it J}-band extinctions of
$A_{eff}=0.35$ and $A_{eff}=0.24$, respectively, which match the measured
extinctions in Table 2.
These results are in agreement with those of our two-component model, as they 
attribute the observed difference between the spiral arm and interarm 
extinction curves to the interarm region having a higher degree of dust 
patchiness (density ratio between high-density and low-density phases).
   
The fits are not perfect (especially in the spiral arm region), but this is 
not unexpected because the full parameter space of the model has not been 
explored.  Future clumpy dust models, suggested by \cite{wit96}, 
that examine the case of an extended star distribution will be particularly 
well suited to explain the behavior of our observed extinction curves. 

\subsection{The Column Depth of Hydrogen in NGC 2207}

Besides flattening the extinction law, an unresolved patchy dust distribution will
cause us to underestimate the total column depth of hydrogen in a galaxy if 
we use the visual extinction for this purpose.  To investigate this we use
the two-component model discussed in sec 4.2.  According to this model, the ISM
is composed of two components, a high density one and a low density one, having
{\it J}-band optical depths of $\tau_1$ and $\tau_2$, respectively.  The
effective extinction through this ISM is given by equation (5), where $f$ is
the area filling factor of high density clumps.  The optical depth that 
corresponds to the total column depth of hydrogen, however, is given by
\begin{equation}
\tau_{H} = f\tau_1 + (1-f)\tau_2
\end{equation}
where $\tau_{H}$ is directly proportional to the column depth of hydrogen,
$N_H$.  The relation between $N_H$ and the {\it V}-band extinction $A_{\it V}$
is given by (\cite{sav79}) as $N_H \approx 1.87 \times 10^{21} $atom cm$^{-2} 
A_{\it V}$.

Adopting the values of $\tau_1$, $\tau_2$, and $f$ shown in Figure 4, and
assuming an $R_{\it V} = 3.1$ law, we find that the total column depth of 
hydrogen in NGC 2207 is: $3.1 \times 10^{21}$ atoms cm$^{-2}$ in the spiral
arm region, and $2.7 \times 10^{21} $atoms cm$^{-2}$ in the interarm region.
If we had used the observed {\it V}-band extinctions (listed in Table 2) to 
estimate $N_H$ we would have found a total column depth equal to $1.8 \times 
10^{21} $atoms cm$^{-2}$ and $1.0 \times 10^{21} $atoms cm$^{-2}$ in the spiral 
arm and interarm regions, respectively.  If, on the other hand, we had used the 
observed {\it K}-band extinctions to estimate $N_H$, we would have found
a total hydrogen column depth equal to $3.8 \times 10^{21} $atoms cm$^{-2}$
and $2.4 \times 10^{21} $atoms cm$^{-2}$ in the spiral arm and interarm regions,
respectively.  In other words, when using the {\it V}-band extinction, we
underestimate the total column depth of hydrogen by a factor of $\sim 2$
solely as a result of an unresolved clumpy dust distribution.  In contrast,
the {\it K}-band extinction provides column depth estimates that are much
more reliable.

The above example clearly demonstrates that it is much better to use the infrared
extinction in order to estimate the total column depth of hydrogen in a galaxy.
The reason for this is that most of the hydrogen is in dense components
which are more easily seen in the infrared than in the visual.

\section {SUMMARY \& DISCUSSION}

The conclusions of this paper are as follows:

\begin{enumerate}
\item The dust in NGC 2207 is mainly concentrated in its spiral arms, leaving 
its interarm regions mostly transparent.  This confirms the results of
\cite{whi96} for other overlapping galaxy pairs.

\item The extinction curve in the spiral arm region is flatter than expected,
as it resembles a Milky Way $R_V=5.0$ dust model.  This confirms the suspicions
of \cite{whi96}.

\item The extinction curve in the interarm region is significantly flatter 
than that of the spiral arm region.  This is, perhaps, the most interesting 
of our results and could be a valuable clue to the nature of the dust 
distribution in different regions of galaxies.  It would be very interesting
to know whether this effect is observed in other spiral galaxies.  Similar 
studies should be performed on other overlapping galaxy pairs in order to 
answer this question.

\item Scattering effects are negligible in this overlapping galaxy pair.

\item An unresolved patchy dust distribution in NGC 2207 is capable of 
producing the extinction curves that we observe.  This confirms the 
suspicions of \cite{wit96} and \cite{whi96}.  Since we are averaging over
boxes which are $\sim 600 pc$ on a side, we expect clumping of dust to be 
significant on these size scales.  Studies that probe the ISM on much smaller 
scales might be less susceptible to patchy dust effects.  

\item Fits of two patchy dust models to the data suggest that the arm-interarm 
difference in the observed extinction law can be explained if the galaxy's 
interarm region has a higher degree of dust patchiness (density ratio between 
high-density and low-density phases) than its spiral arm region.

\item It is clear that unresolved clumping of dust will cause one to 
underestimate the average column depth of gas in a galaxy if based solely on 
the visual extinction.  It is much better to use the infrared extinction for 
this purpose.
\end{enumerate}

It would be interesting to compare our derived extinction laws for NGC 2207 
with those that might be obtained from the galaxy's colors (e.g., \cite{rix93} 
use the colors of M51 to determine its optical depth).  
Also, future efforts should be made to obtain high resolution visual and 
infrared images of very distant overlapping galaxies.
The method developed by \cite{whi96}, which makes use of overlapping galaxy
pairs, is a valuable tool for probing the ISM in distant galaxies.  It 
is, perhaps, one of the few ways to study the dust content in high redshift 
galaxies.

\acknowledgments

We acknowledge helpful discussions and correspondence with D. L. DePoy,
J. A. Frogel, B. S. Gaudi, A. Gould, W. C. Keel, D. M. Terndrup and A. N. Witt.  
We thank A. N. Witt and K. Gordon for sending us their model results in 
electronic format.  We thank the referee for his/her helpful comments.  
We are grateful to G. Tiede for helping obtain the data.
We thank the CTIO TAC for generous allocations of time for the Galaxy 
Survey project.  The OSU Galaxy Survey is being supported in part by 
NSF grant AST 92-17716.

\newpage
\begin{center}
\centerline{\small TABLE 1. Derived Intrinsic Galaxy Colors}
\smallskip
\begin{tabular}{lcclcl}
\hline
\hline
\multicolumn{1}{c}{color} &
\multicolumn{1}{c}{ } &
\multicolumn{2}{l}{NGC 2207 } &
\multicolumn{2}{l}{IC 2163 } \\
\multicolumn{1}{c}{ } &
\multicolumn{1}{c}{ } &
\multicolumn{1}{c}{arm} &
\multicolumn{1}{c}{interarm} &
\multicolumn{1}{c}{arm} &
\multicolumn{1}{c}{interarm} \\
\hline
B-V & & 0.52 & 0.71 & 0.80 & 0.66 \\
V-R & & 0.45 & 0.52 & 0.63 & 0.56 \\
R-J & & 1.25 & 1.30 & 1.80 & 1.64 \\
J-H & & 0.77 & 0.67 & 0.66 & 0.63 \\
H-K & & 0.07 & 0.29 & 0.30 & 0.27 \\
\hline
\hline
\end{tabular}
\end{center}
\noindent
Note---Estimated errors range from 0.04 mag in {\it K} to 0.10 mag in {\it B}.

\clearpage
 \begin{center}
\centerline{\small TABLE 2. The Extinction through NGC 2207}
\smallskip
\begin{tabular}{lllc}
\hline
\hline
\multicolumn{1}{l}{ } &
\multicolumn{1}{c}{Spiral arm \ \ \ \ } &
\multicolumn{1}{c}{Interarm \ \ \ \ \ } &
\multicolumn{1}{c}{ } \\
\multicolumn{1}{c}{Filter} &
\multicolumn{1}{l}{ $A_{\lambda}$ } &
\multicolumn{1}{l}{ $A_{\lambda}$ } &
\multicolumn{1}{c}{F/B} \\
\hline
B & 1.32 $\pm0.10$ & 0.60 $\pm0.10$ & 0.27\\
V & 0.96 $\pm0.10$ & 0.52 $\pm0.10$ & 0.21\\
R & 0.80 $\pm0.09$ & 0.44 $\pm0.09$ & 0.17\\
J & 0.35 $\pm0.06$ & 0.24 $\pm0.07$ & 0.10\\
H & 0.27 $\pm0.05$ & 0.19 $\pm0.06$ & 0.11\\
K & 0.22 $\pm0.04$ & 0.14 $\pm0.06$ & 0.09\\
\hline
\hline
\end{tabular}
\end{center}
\noindent
Notes---$A_{\lambda}$ is the extinction measured at a wavelength $\lambda$.  F/B
is the ratio of the foreground galaxy surface brightness to that of the background
galaxy.  Uncertainties shown are photometric uncertainties and uncertainties due
to asymmetry in the background galaxy added in quadrature.
\newpage

\begin{center}
\centerline{\small TABLE 3. Scattering Estimates}
\smallskip
\begin{tabular}{llllr}
\hline
\hline
\multicolumn{1}{l}{Filter} &
\multicolumn{1}{l}{$B_{tot}$} &
\multicolumn{1}{l}{$S_{max}$} &
\multicolumn{1}{l}{$\frac{S_{max}}{F + Be^{-\tau}}$} & 
\multicolumn{1}{r}{$\frac{\Delta A_{\lambda}}{A_{\lambda}}$} \\
\multicolumn{1}{l}{} &
\multicolumn{1}{l}{(mag)} &
\multicolumn{1}{l}{(mag/$\Box''$)} &
\multicolumn{1}{l}{} &
\multicolumn{1}{r}{} \\
\hline
B & 13.03 & 25.49 & 4.0\% & 3.3\% \\
V & 12.28 & 25.07 & 2.4\% & 2.7\% \\
R & 11.70 & 24.71 & 1.7\% & 2.4\% \\
J & 10.00 & 23.90 & 0.5\% & 1.7\% \\
H &  9.35 & 23.52 & 0.4\% & 1.6\% \\
K &  9.10 & 23.51 & 0.3\% & 1.5\% \\
\hline
\hline
\end{tabular}
\end{center}
\noindent
Notes---$B_{tot}$ is the total brightness of the background galaxy; $S_{max}$ is
the maximum surface brightness that could be scattered into the line of sight;
$F$ and $B$ are the surface brightnesses of the foreground and background galaxies,
respectively; $\tau$ is the extinction optical depth through the disk of the 
foreground galaxy; $A_{\lambda}$ is the extinction of light (as a function of 
wavelength) through the disk of the foreground galaxy; $\Delta A_{\lambda}$ is
the maximum change in our extinction estimates that could result from scattering
of light.

\begin{figure}
\plotfiddle{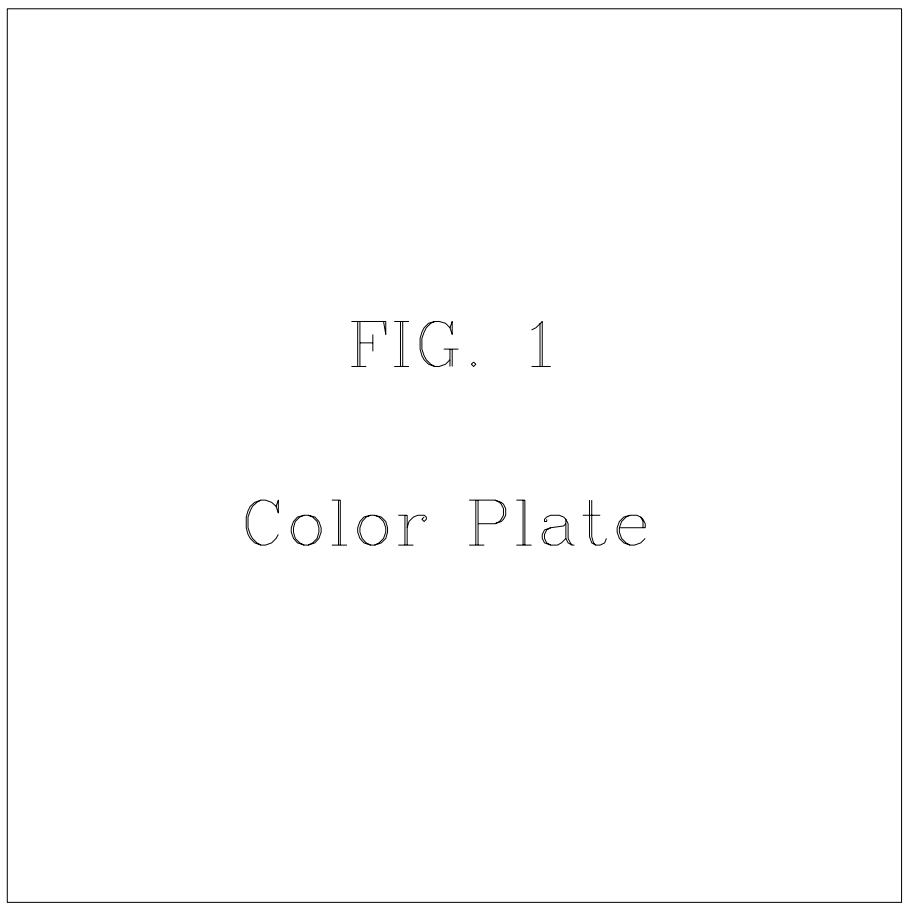}{5.0truein}{0}{70}{70}{-225}{-90}
\caption{Color image of NGC 2207 (right) and IC 2163 (left) constructed 
from {\it HR} and {\it B}-band images.  The image has dimensions ($6.\hskip-2pt'9 
\times 8.\hskip-2pt'6$).  Note that IC 2163 can be easily seen through NGC 2207.  
Also note the reddening of background galaxy light in the foreground galaxy's 
spiral arm region.}
\end{figure} 

\begin{figure}
\plotfiddle{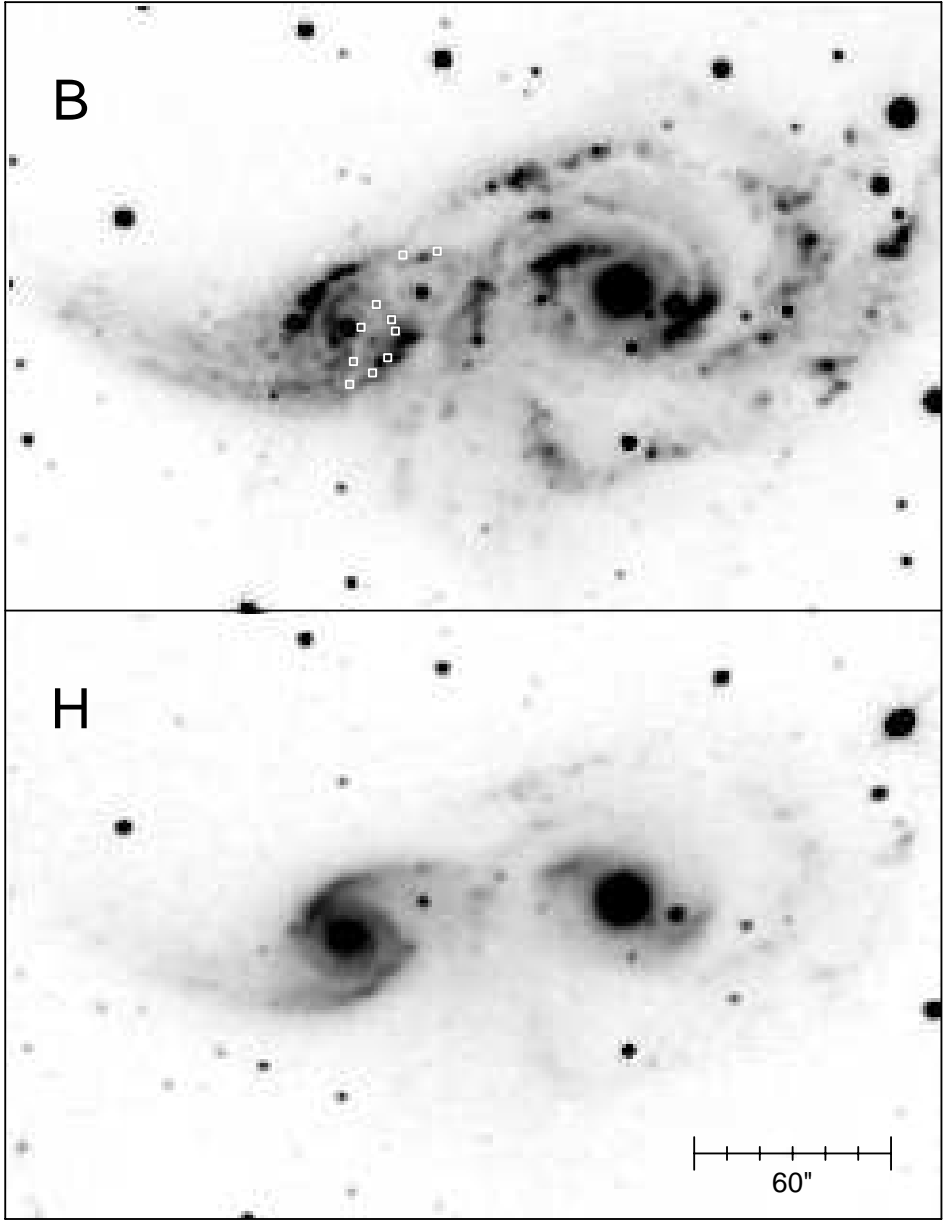}{5.0truein}{0}{70}{70}{-225}{-90}
\vskip 38pt
\caption{{\it B} and {\it H} images of NGC 2207 (right) and IC 2163 
(left).  In these images North is up and East is to the left.
The square regions in which we measure the extinction are marked on the 
{\it B} image (there are five in the spiral arm and five in the interarm 
region).  The regions are approximately $3.\hskip-2pt''5$ wide ($1''= 170$pc 
for a distance to the galaxy of 35Mpc).  The angular scale of the images is 
given at the bottom.  Note that the {\it B} image shows that the foreground 
dust lane in NGC 2207 causes significant extinction of the background galaxy light, 
whereas in the {\it H} image the extinction is minimal.}
\end{figure}

\begin{figure}
\plotfiddle{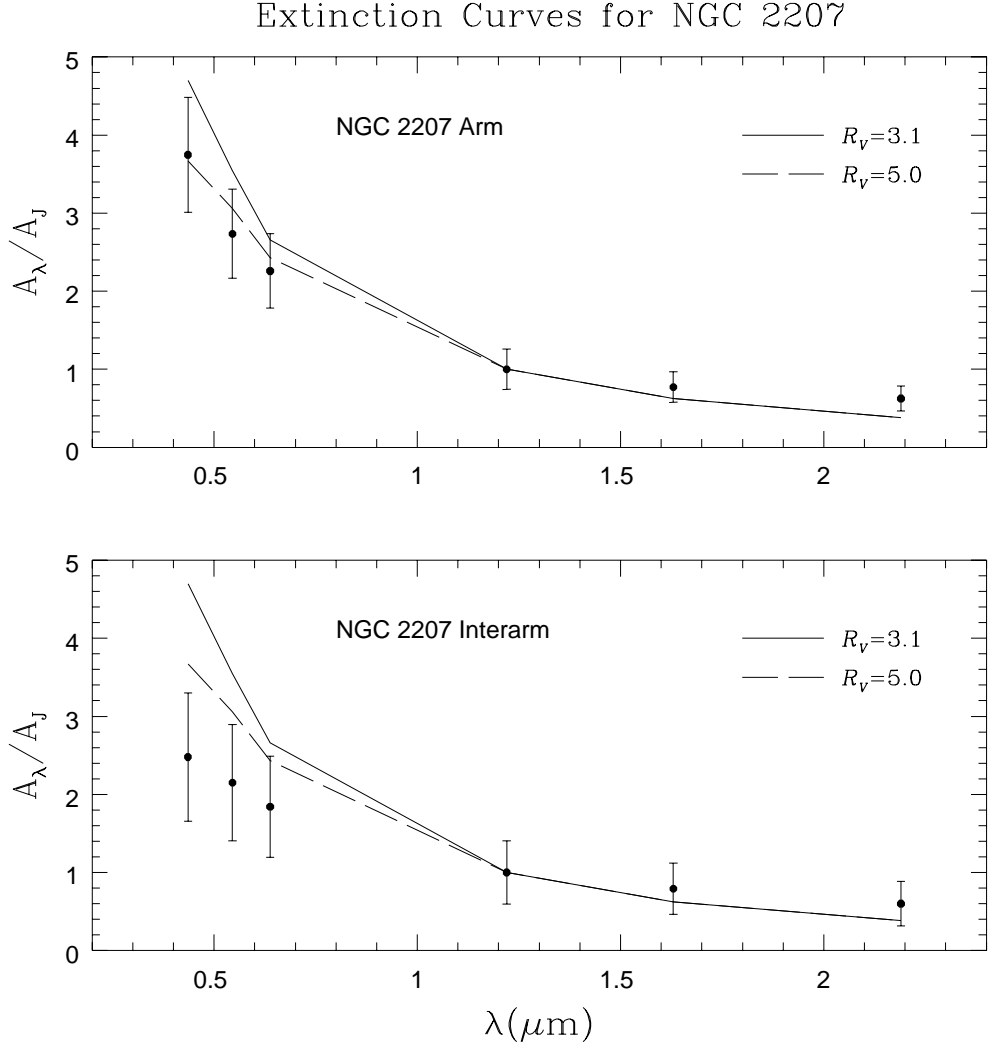}{5.0truein}{0}{70}{70}{-225}{-100}
\caption{Extinction vs. wavelength for the dust content in NGC 2207's 
spiral arm (top) and interarm (bottom) regions.  The points represent 
our extinction measurements $A_{\lambda}$, as a function of wavelength $\lambda$,
at {\it BVRJHK}, normalized to $A_{\it J}$, the 
extinction in {\it J}, along with their $1\sigma$ uncertainties (a discussion 
on errors can be found in sec. 3.2).  The solid and dashed lines 
represent $R_V=3.1$ and $R_V=5.0$ Galactic extinction laws respectively (where
$R_{\it V} = A_{\it V}/E({\it B}-{\it V})$).  
Note that the extinction laws in both the spiral arm and interarm regions of
NGC 2207 are flatter (``greyer'') than a standard $R=3.1$
Galactic extinction law.  Also note that the interarm extinction law 
is substantially flatter (``greyer'') than the spiral arm extinction 
law.}
\end{figure}

\begin{figure}
\plotfiddle{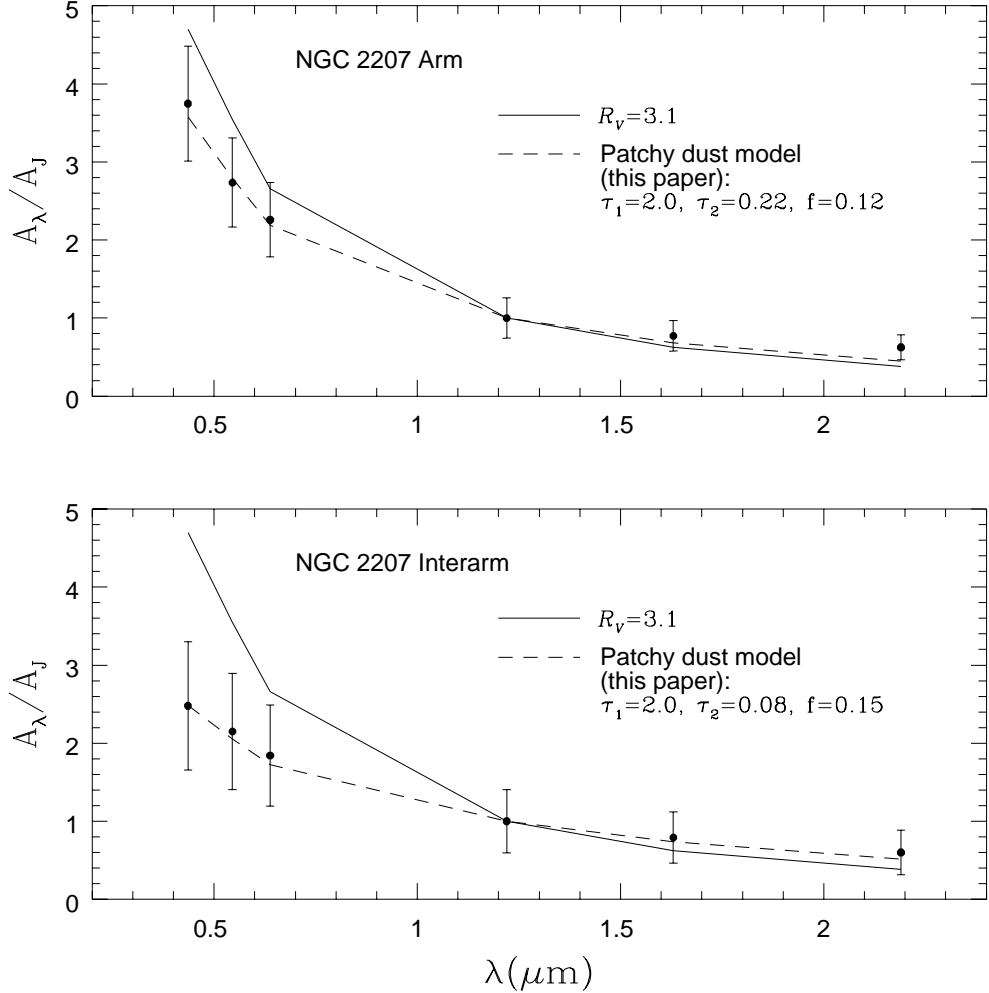}{5.0truein}{0}{70}{70}{-225}{-120}
\vskip 24pt
\caption{Effective extinction laws for a two-component dust
model.  The two panels show the effective extinction curves tailored to
fit the spiral arm (top) and interarm (bottom) extinction curves. 
The points represent our extinction measurements $A_{\lambda}$, as a function of 
wavelength $\lambda$, at {\it BVRJHK}, normalized to $A_{\it J}$, the 
extinction in {\it J}, along with their $1\sigma$ uncertainties.  The solid lines 
represent an $R_V=3.1$ Galactic extinction law (where $R_{\it V} =
A_{\it V}/E({\it B}-{\it V})$).
The dashed lines represent the effective extinction curves which result from 
our patchy dust model.  The values of $\tau_1$, $\tau_2$ (the {\it J}-band optical 
depths of high and low density dust components, respectively), and $f$ (the area 
filling factor of high density clumps) corresponding to these curves are shown.  
These parameters are defined in sec. 4.2.  Note that the two-component 
model does an excellent job in reproducing our observed extinction curves 
(this figure is meant to illustrate how a patchy dust distribution is capable 
of explaining our observed extinction curves.  The specific models shown above 
are not unique and should not be assigned much physical meaning.) }
\end{figure}

\begin{figure}
\plotfiddle{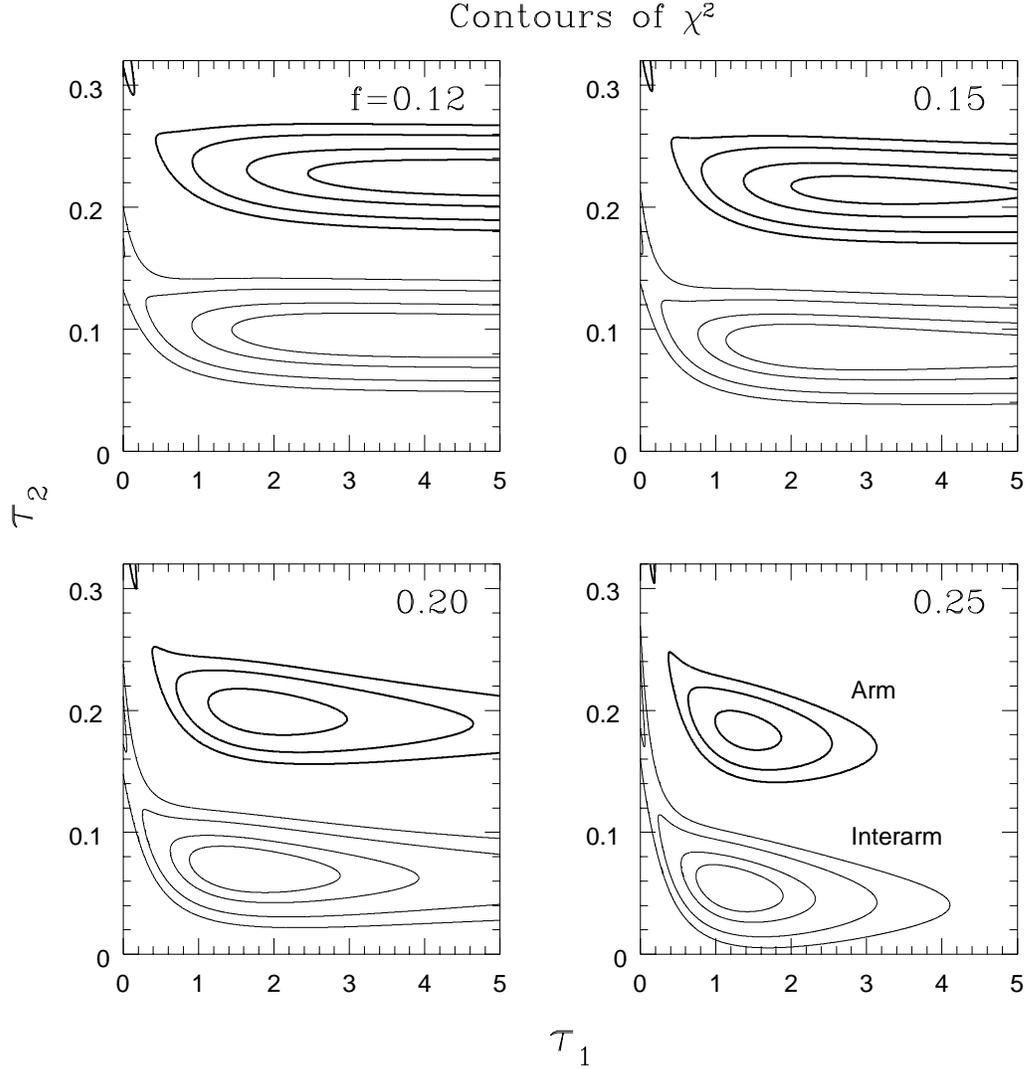}{5.0truein}{0}{70}{70}{-225}{-120}
\vskip 24pt
\caption{Contours of constant $\chi^2$ for our two-component dust
model.  The four panels show contours of constant reduced-$\chi^2$ as a functionof the paramet
ers $\tau_1$ and $\tau_2$ (the {\it J}-band optical 
depths of high and low density dust components, respectively).  The contours 
represent fits of this model to our observed spiral arm and interarm 
extinctions for area filling factors of: $f=0.12$ (top left), $f=0.15$ (top right), 
$f=0.20$ (bottom left), and $f=0.25$ (bottom right). 
Bold and light contours represent fits to the spiral arm and interarm extinction 
curves, respectively.  The contours shown are for values of the reduced-$\chi^2$equal to 3, 2,
 1, and 0.5 (going from outer to inner contours).  Where only three 
contours are shown, these are for values of the reduced-$\chi^2$ equal to 3, 2, and
1.  We note that $\tau_2$ is reasonably tightly constrained and does not show a great dependen
ce on $\tau_1$ or $f$.  $\tau_1$, on the other hand, is not tightly 
constrained.  These ``golf club'' contours reveal the main difference between the 
spiral arm and interarm regions: the low density dust optical depth, $\tau_2$, 
is smaller in the interarm region than it is in the spiral arm region. }
\end{figure}

\begin{figure}
\plotfiddle{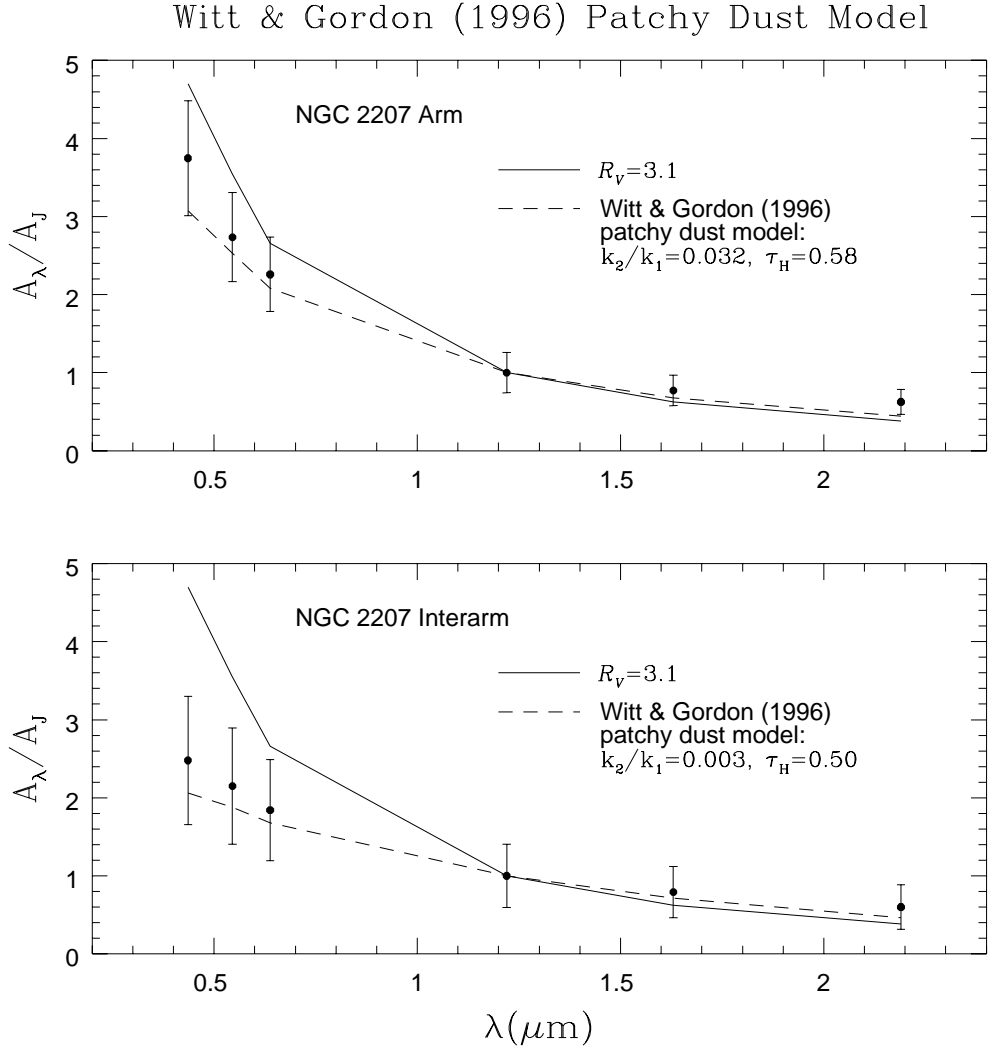}{5.0truein}{0}{70}{70}{-225}{-120}
\vskip 24pt
\caption{Effective extinction laws for the Witt \& Gordon (1996)
clumpy dust model.  The two panels show the best-fit effective extinction curves 
for the spiral arm (top) and interarm (bottom) extinction curves.  The points 
represent our extinction measurements $A_{\lambda}$, as a function of 
wavelength $\lambda$, at {\it BVRJHK}, normalized to $A_{\it J}$, the 
extinction in {\it J}, along with their $1\sigma$ uncertainties.  The solid lines 
represent an $R_V=3.1$ Galactic extinction law (where $R_{\it V} =
A_{\it V}/E({\it B}-{\it V})$).
The dashed lines represent the best-fit effective extinction curves which 
result from the Witt \& Gordon (1996) clumpy dust model.  The best-fit values of 
$k_2/k_1$ (the density ratio between low density and high density phases) and 
$\tau_H$ (the {\it J}-band optical depth of the equivalent homogeneous dust 
distribution) corresponding to these curves are shown.  Note that this model 
suggests that the observed difference between the spiral arm and interarm 
extinction curves is caused by the interarm region having a higher density 
contrast between high and low density clouds ($k_1/k_2$) than the spiral arm 
region.}
\end{figure}

\end{document}